\documentclass[iop,revtex4]{emulateapj}
    \pdfoutput=1
  \usepackage{color}
  \usepackage{times}
  \usepackage{graphicx}
  \usepackage{subfigure}
  \usepackage{float}
  \usepackage[  backref,breaklinks,colorlinks,citecolor=blue,linkcolor=blue]{hyperref}
    \usepackage{txfonts}
  \usepackage{epsfig}
  \usepackage{lscape}
  \usepackage{epstopdf}
  \usepackage{CJK}
      \def\co{$^{12}$CO}
  \def\tco{$^{13}$CO}
  \def\ceo{C$^{18}$O}
  
    \renewcommand{\arraystretch}{2.0}
  
  \newcommand{\hii}{H~\textsc{ii}}
  \newcommand{\msun}{$M_\odot$}
  \newcommand{\lsun}{$L_\odot$}
  \newcommand{\kms}{km~s$^{-1}$}
  \newcommand{\hcop}{HCO$^+$}
  \newcommand{\hcopone}{HCO$^+$~$J=1-0$}
  \newcommand{\hcnone}{HCN~$J=1-0$}
  \newcommand{\ceoone}{C$^{18}$O~$J=1-0$}
  \newcommand{\tcoone}{$^{13}$CO~$J=1-0$}
\journalinfo{Accepted by the Astrophysical Journal}
\submitted{Draft Version, \today}
\shorttitle{Dense gas in CO selected cores associated PGCCs}
\shortauthors{Yuan et al.}
\begin{document}
\begin{CJK*}{UTF8}{gbsn}
\title{Dense gas in molecular cores associated with \emph{Planck}  Galactic cold clumps}
\author{Jinghua Yuan (袁敬华)\altaffilmark{1}, Yuefang Wu\altaffilmark{2}\footnotemark[\dag], 
Tie Liu\altaffilmark{3}, Tianwei Zhang\altaffilmark{4}, Jin Zeng Li\altaffilmark{1}, Hong-Li Liu\altaffilmark{1}, Fanyi Meng\altaffilmark{5}, \\
 Ping Chen\altaffilmark{2}, Runjie Hu\altaffilmark{2}, and Ke Wang\altaffilmark{6}}
\affil{$^1$National Astronomical Observatories, Chinese Academy of Sciences, 20A Datun Road, Chaoyang District, Beijing 100012, China;}
\affil{$^2$Department of Astronomy, Peking University, 100871 Beijing, China;}
\affil{$^3$Korea Astronomy and Space Science Institute 776, Daedeokdae-ro, Yuseong-gu, Daejeon, Republic of Korea 305-348;}
\affil{$^4$Peking University Health Science Center, Xueyuan Road 38th, Haidian District, Beijing 100191, China;}
\affil{$^5$Physikalisches Institut, Universit\"{a}t zu K\"{o}ln, Z\"{u}lpicher Str. 77, 50937, Germany;} 
\affil{$^6$European Southern Observatory, Karl-Schwarzschild-Str. 2, D-85748 Garching bei M\"{u}nchen, Germany}
\footnotetext[\dag]{ywu@pku.edu.cn}

\begin{abstract}

    We present the first survey of dense gas towards \emph{Planck} Galactic Cold Clumps (PGCCs). 
    Observations in the $J=1-0$ transitions of \hcop~and HCN towards 621 molecular 
    cores associated with PGCCs were performed using the Purple Mountain 
    Observatory 13.7-m telescope. Among them, 250 sources have detection, 
    including 230 cores detected in \hcop~and 158 in HCN. Spectra of the $J=1-0$ 
    transitions from \co, \tco, and \ceo~at the centers of the 250 cores were 
    extracted from previous mapping observations to construct 
    a multi-line data set. The significantly low detection rate of asymmetric 
    double-peaked profiles, together with the well consistence among central velocities 
    of CO, \hcop, and HCN spectra, suggests that the CO-selected \emph{Planck} cores 
    are more quiescent compared to classical star-forming regions. 
    The small difference between line widths of \ceo~and HCN indicates
    that the inner regions of CO-selected \emph{Planck} cores are not more turbulent 
    than the exterior. The velocity-integrated intensities and 
    abundances of \hcop~are positively correlated with those of HCN, 
    suggesting these two species are well coupled and chemically connected. 
    The detected abundances of both \hcop~and HCN are significantly 
    lower than values in other low- to high-mass star-forming regions. 
    The low abundances may be due to beam dilution. On the basis of the inspection of 
    the parameters given in the PGCC catalog, we suggest that there may be about 
    1 000 PGCC objects having sufficient reservoir of dense gas to form stars.

\end{abstract}

\section{Introduction}\label{s-intro}

   Stars form in dense regions of molecular clouds. However, some physical 
   and chemical properties of cold compact objects that breed stars are sill 
   poorly understood. A key approach would be to statistically investigate 
   cold dense clumps in the Galaxy. In the last decade, numerous dense clumps 
   revealed by surveys in continuum at (sub-)millimeter domain using 
   ground-based facilities and the \emph{Herschel Space Observatory} 
   \citep[e.g.][]{sch09,and10,mol10,agu11} have led to significant advancement. 
   However, these surveys are mainly confined in the Galactic plane and 
   several well-known nearby star-forming regions, an all-sky survey at 
   multi-bands is crucial to the understanding of star formation in 
   different environments.

   The \emph{Planck} satellite \citep{tau10,pla11a} carried out the first
   all-sky survey at multi-bands in the submm-to-mm range with
   unprecedented sensitivity and provides an inventory of cold condensations
   of interstellar matter in the Galaxy. The first all-sky Cold Clump Catalogue
   of \emph{Planck} Objects (C3PO) released in \citet{pla11d} consists of 10,342
   cold sources that stand out against a warmer environment. The C3PO clumps are
   cold with dust temperatures ranging from 7 K to 19 K, with a distribution
   peaking around 13 K. A detailed analysis of ten C3PO clumps carried out by \citet{pla11c} 
   revealed low column densities of
    {$N_\mathrm{H_2}\sim(0.1-1.6)\times10^{22}$ cm$^{-2}$} which can only be treated as lower limits
   due to the poor resolution of \emph{Planck} bands. Among the C3PO clumps,
   915 early cold cores (ECCs) were identified based on criteria of $S/N>15$ and
   $T<14$ K where $S/N$ is the signal-to-noise ratio. The ECC sample is delivered as
   a part of the $Planck$ Early Release Compact Source Catalogue \citep[ERCSC,][]{pla11b}. 
   As the high-S/N and cold tail of the
   C3PO objects, ECCs are excellent targets for studying the earliest stages
   of star formation \citep{juv10,juv11,juv12,juv15,par15}.
   Recently, \citet{pla15} released the \emph{Planck} Catalog of Galactic
   Cold Clumps (PGCCs). As the full version of the ECC catalog, the PGCC 
   catalog provides 13 188 Galactic sources spreading across the whole
   sky. Containing sources in very different environments, the PGCC
   catalog is useful for investigating the evolution from molecular 
   clouds to cores.

   To reveal the molecular characteristics, we carried out
   survey observations in the $J=1-0$ transitions of \co, \tco, and \ceo~towards
   more than 600 $Planck$ cold clumps selected from the ECC catalog in both
   single-pointing and on-the-fly modes \citep{wu12,liu12,liu13,liu15,men13}. We found that
   $Planck$ cold clumps have the smallest line widths compared to other star-formaing
   samples \citep[e.g., IRDCs, ][]{wan09,wan14} indicating quiescent features and nature of very early
   evolutionary stages. 

    Although some kinematic properties have been revealed, the low critical densities
    of CO and its isotopologues prevent us investigating the denser inner regions. 
    Observations in dense gas tracers are crucial to
    unveiling the physical and chemical features of the interior of these objects.

    In this paper, we report on a survey of the $J=1-0$ transitions of \hcop~and HCN
    towards the  {smaller sub-structures enclosed in the ECCs 
    which have been previously mapped in CO (see section \ref{s-obs})}. 
    Combining spectra of CO and its isotopologues extracted 
    from previous mapping observations, we have carried out a comprehensive study
    of dense gas in these sources.
    This paper is arranged as follows. We present a description of the sample
    and observations in Section 2 and the results in Section 3.
    In Section 4, we try to comprehensively
    discuss the data. The main findings are summarized in Section 5.

    \begin{figure*}
    \centering
    \includegraphics[width=0.9\textwidth]{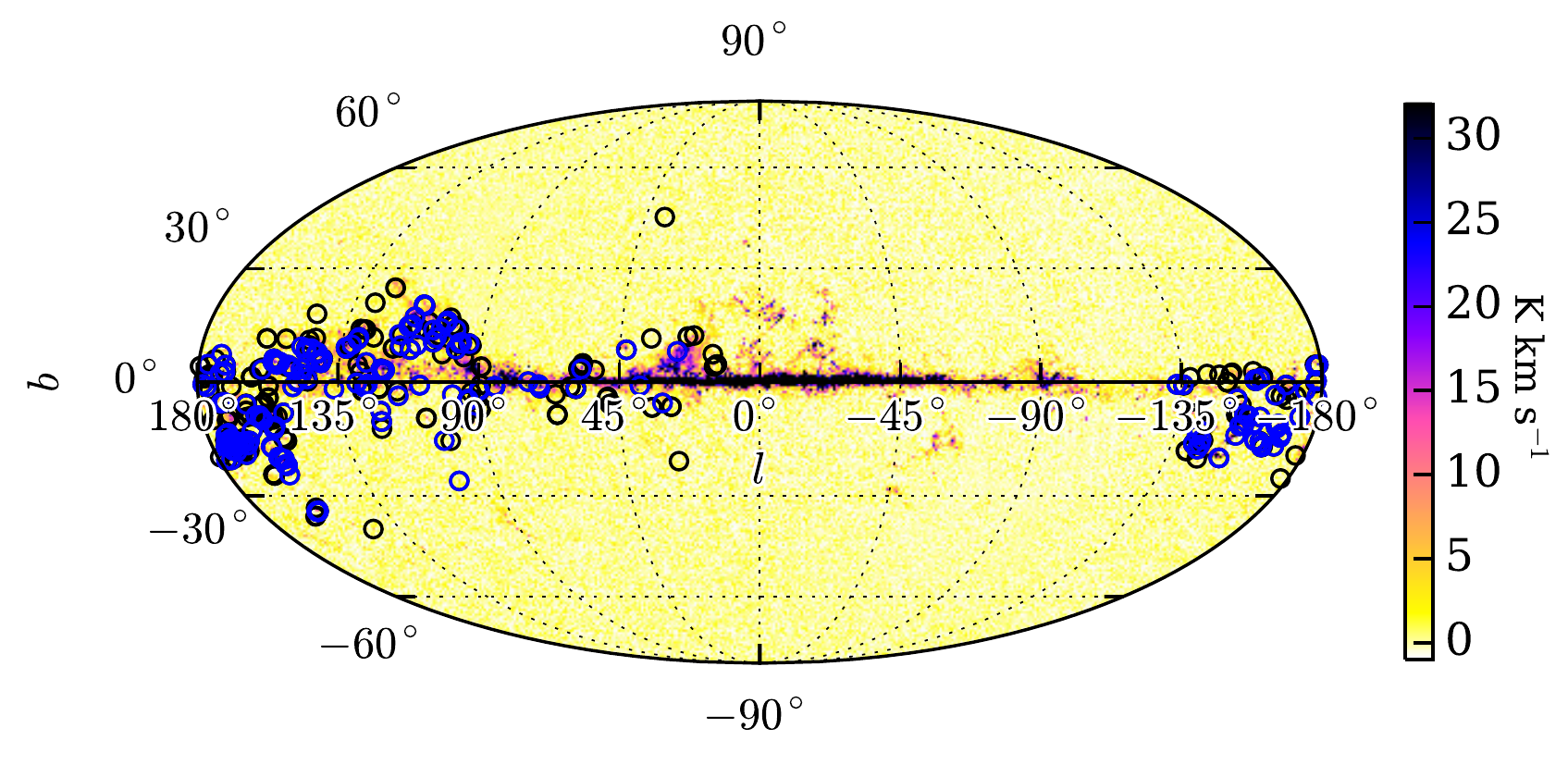}
    \caption{Spatial distribution of the observed sources (black circles) and 
    the ones with valid detection (blue circles). The background shows CO $J=1-0$ 
    emission of the Type 3 map from \cite{pla14}.}\label{fig:sour-on-CO}
    \end{figure*}

\section{The Sample and Observations}\label{s-obs}

       {
      As mentioned in section \ref{s-intro}, more than 600 $Planck$ cold clumps have been 
      mapped in the $J=1-0$ transitions of \co, \tco, and \ceo. A $22'\times22'$ region
      for each clump was mapped with a spatial resolution of about $52''$. 
      Only the central $14'\times14'$ portion, where the sensitivity is
      high enough, have been taken into account. 
      Details about the mapping observations are referred to \citet{liu12} and 
      \citet{men13}. 
      As demonstrated in \citet{men13}, we have identified sub-structures in regions 
      defined by a contour of 60\% of the peak intensity of \tco. In total, 621 smaller
      substructures have been identified \citep[][T. Zhang et al. 
      in prep; P. Chen et al. in prep; R. Hu et al. in prep]{liu12,men13}, and are referred as 
      CO-selected cores in this work to be distinguished from the parental clumps. 
      These CO-selected cores have
      source-averaged column densities ranging from $7.6\times10^{20}$ cm$^{-2}$
      to $3.7\times10^{22}$ cm$^{-2}$ with a mean value of $7.2\times10^{21}$ cm$^{-2}$. The
      source-averaged volume densities are in the range of $(0.1-68.6)\times10^3$ cm$^{-3}$ 
      with a mean of $4.6\times10^3$ cm$^{-3}$.
      The low mean excitation temperature of CO ($\sim11$ K) suggests quiescent
      features. The distances to 164 CO-selected cores have been adopted 
      from the PGCC catalog \citep{pla15}. The distances to other 382 sources, whose distances are not 
      provided in the PGCC catalog, are from \citet{wu12}.
      Note that we have confined the Galactocentric distance to be smaller than 20 kpc while
      matching the sample with the catalogs of \citet{wu12} and \citet{pla15}. The distances range
      from 0.1 to 14.7 kpc with a mean value of 1.3 kpc.
      }

          Single-pointing observations of the CO-selected cores in \hcopone\ (89.189 GHz) and \hcnone\ 
      (88.632 GHz) were carried out using the Purple Mountain Observatory (PMO) 13.7-m telescope with 
      the position-switch mode from June to July of 2013 and from May to June of 2014. The half power 
      beam width (HPBW) and main beam efficiency  at 89 GHz are about 59\arcsec\ and 55\%, 
      respectively. The pointing and tracking accuracies were both better than 5\arcsec\ 
      determined by scanning solar planets. A double sideband SIS receiver was used to cover 
      both lines in the upper sideband (USB) which has a width of 1 GHz allocated to 16384 
      channels \citep{sha12}. The spectral resolution is about 61 kHz, corresponding to a 
      velocity resolution of 0.21 \kms. In our observations, the off position for each source 
      was originally set to be 30\arcmin\ west away from the target along the right ascension direction. 
       {We have carefully checked the the off positions with direct observations to
      find that most of them are emission-free in both \hcop~and HCN. For the five sources, which 
      show anomalous absorption features in the spectra indicating possible line contamination, the
      off positions were reset to the opposite direction, i.e. 30\arcmin\ east from the target.}
      The on-source time was about 5 minutes for most sources, and longer than 10 minutes for 
      a small number of objects observed in 2013. The system temperatures are in the range of 
      133 to 363 K, with a mean value of 190 K. The CLASS software of the GILDAS package 
      \citep{pet05} was used to reduce the data. In order to achieve a higher 
      sensitivity, we smoothed the spectra to a lower velocity resolution of 0.42 \kms. 
      The final reduced spectra have $1\sigma$ noises ranging from 0.02 to 0.23 K, 
      with a mean value of 0.07 K  {and a median of 0.06 K.}

\section{Results}\label{s-results}

\begin{figure}
    \centering
    \includegraphics[width=0.45\textwidth]{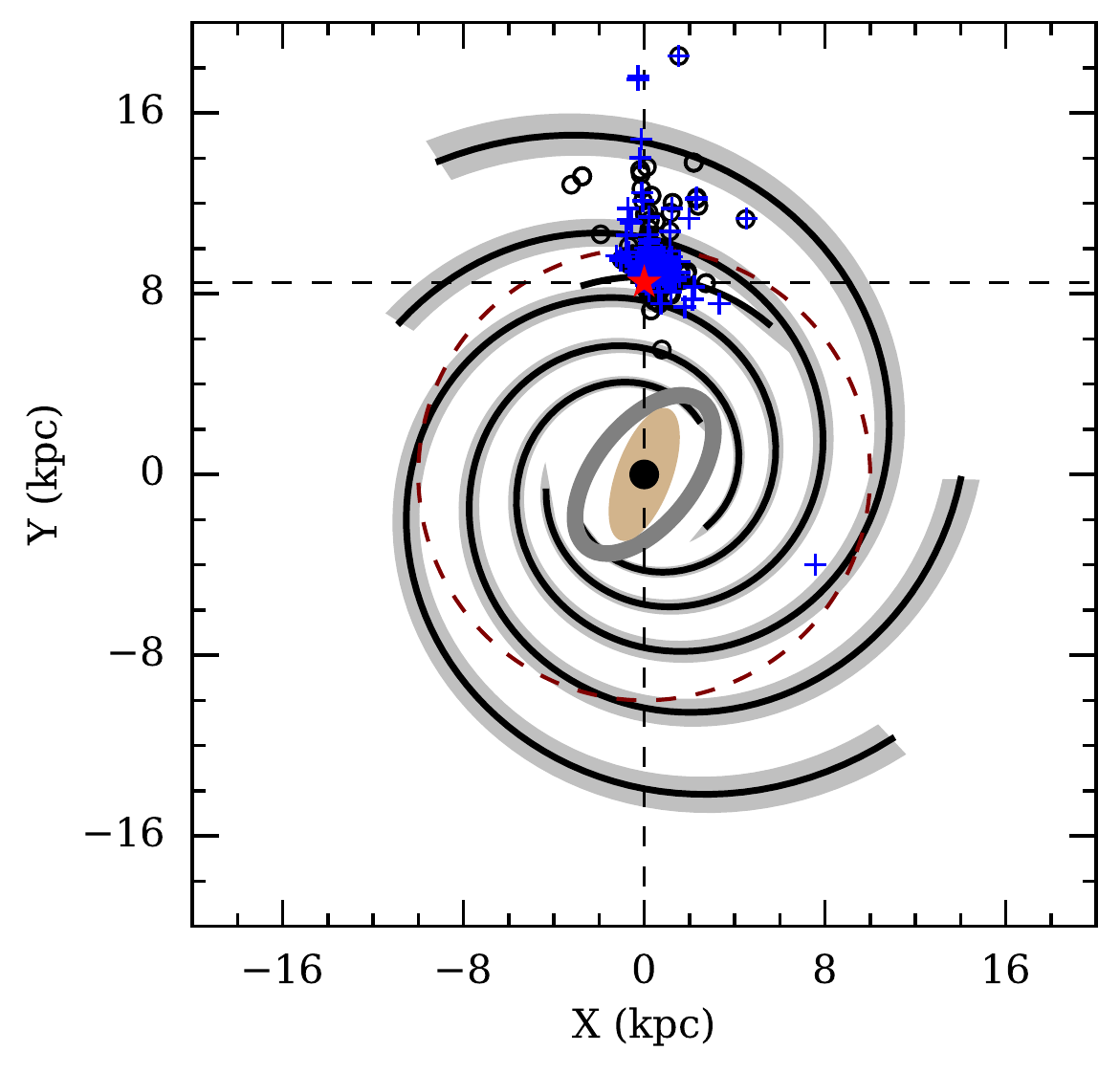}
    \caption{Spatial distribution of the sources with (blue crosses) and without (black circles) 
    detection in the 
    face-on Milky Way. The spiral arms are reproduced using the 4-arm model of \citet{hou14} 
    with $R_0=8.5$ kpc and $\Theta_0=220$ \kms~fitted to \hii~regions. The large dashed circle 
    has a radius of 10 kpc.
    }\label{fig:mw-faceon}
    \end{figure}

\begin{figure*}
    \centering
    \includegraphics[width=0.9\textwidth]{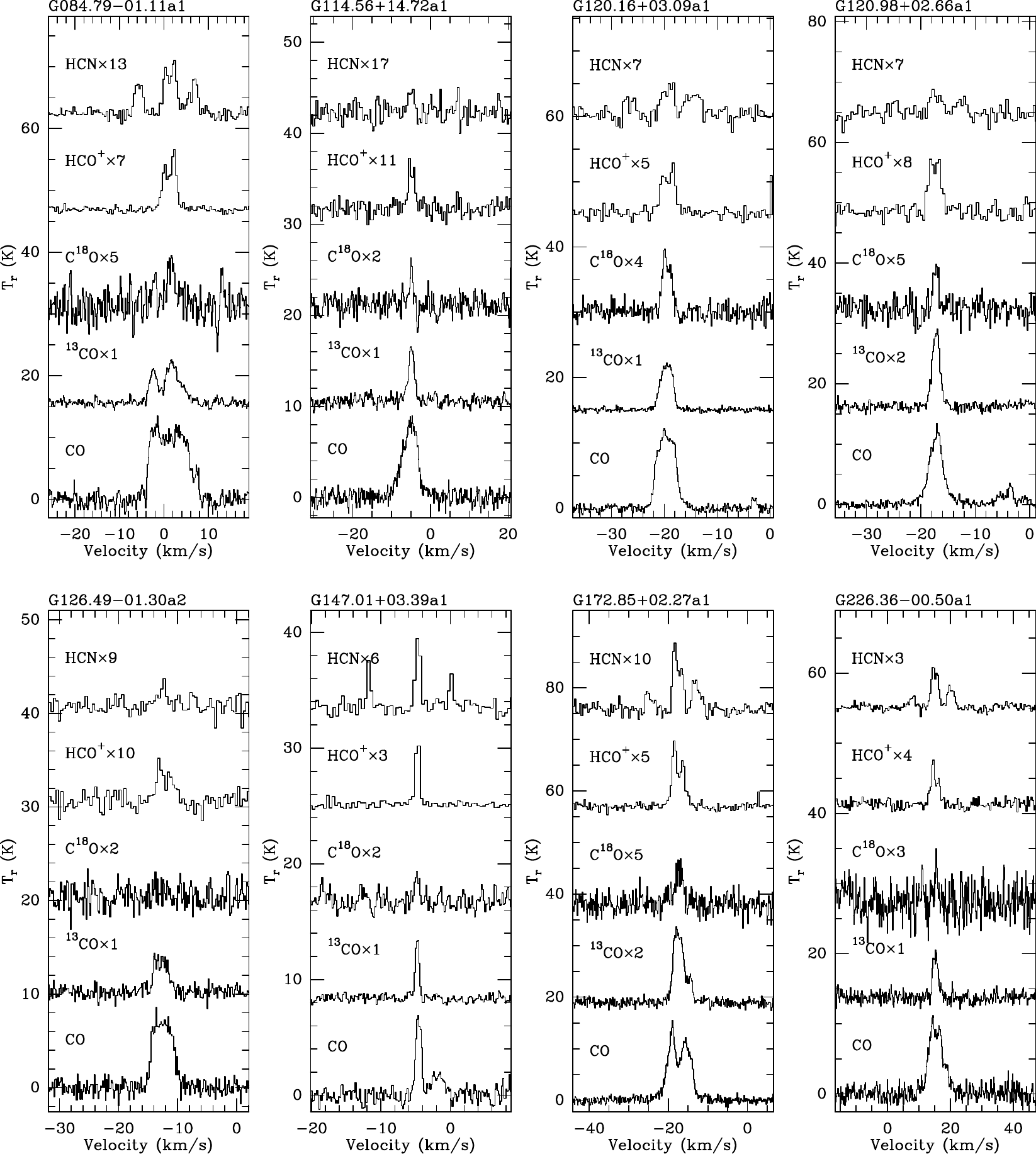}
    \caption{Example spectra.}\label{fig:spectra}
    \end{figure*}

    Among the 621 observed molecular cores, 250 ones have valid detection in either \hcopone\ or \hcnone, including 138 cores detected in both lines, 92 cores only detected in \hcopone, and 20 cores only detected in \hcnone. The designations and coordinates of these 250 sources are given in the first three columns of \autoref{tb-obs-dense}. In \autoref{fig:sour-on-CO}, they are represented with
    filled circles. We note that the spatial distribution of CO-selected
    cores is biased towards nearby star-forming regions (e.g., Perseus, Taurus and Orion) while the Galactic plane is under-represented. This is because of the higher confusion in the Galactic plane, so that the S/N tends to be lower there, but also because the dust temperatures in the plane tend to be higher conflicting to the selection criteria of ECCs \citep{pla11d}.  {The distribution of the targets in the face-on Milky Way is
    shown in \autoref{fig:mw-faceon}. Most sources ($\sim85\%$) reside in the 10 kpc inner region from the Galactic center 
    with a concentration near the Sun.} Example spectra of several sources
    are presented in \autoref{fig:spectra}. Most sources have one velocity component
    except for four ones which show two velocity components. In all the 250 sources, 254 velocity components have been detected.

    \subsection{Observed Parameters}\label{sec:obs-para}

    From previous mapping observations, we have extracted spectra of the $J=1-0$ transitions of \co, \tco\ and \ceo\ at the peak of each core to construct a data set with five lines. For spectra of \co, \tco, \ceo\ and \hcop, peak intensities, centroid velocities, and line widths of all the 250 cores have been obtained by fitting Gaussian profiles.

     {For \hcnone, we have performed hyperfine structure fitting by initiating the 
    intensity ratios of $I_{F=1-1}/I_{F=2-1}=0.6$ and $I_{F=0-1}/I_{F=2-1}=0.2$~using 
    the HFS method in CLASS which resulted in four parameters \emph{p1, p2, p3}, and \emph{p4}. Here, 
    \emph{p2} and \emph{p3} are the centroid velocity and line width. The intensities and 
    optical depths of the main and satellite components can be expressed as functions of
    \emph{p1} and \emph{p4}, }
    \begin{eqnarray}
    T_\mathrm{main} &=& \frac{p1}{p4}\left(1-e^{-p4}\right), \nonumber \\
    T_\mathrm{satellite} &=& \frac{p1}{p4}\left(1-e^{-R\cdot p4}\right),  \\
    \tau_\mathrm{main} &=& p4,  \nonumber  \\
    \tau_\mathrm{satellite} &=& R\cdot p4.  \nonumber 
    \end{eqnarray}
     {Here, $R$ is the intrinsic intensity ratio of the satellite component to the main component. 
    The detailed procedure of the hyperfine fitting is referred to the online documents of the CLASS 
    software\footnotemark[1]}. The resulting observational parameters for \hcopone~and \hcnone~are given in \autoref{tb-obs-dense}.

    \footnotetext[1]{\url{http://www.iram.fr/IRAMFR/GILDAS/doc/html/class-html}}

    \subsection{Derived Parameters}\label{sec:derived-para}

    For the sake of revealing physical features of these molecular cores, some fundamental parameters have been deduced from the observed spectra under the assumption of local thermal equilibrium (LTE). The measured brightness temperature ($T_r$) for a specific transition as a function of excitation temperature ($T_\mathrm{ex}$) can be expressed as
    \begin{eqnarray}
    T_{r} &=& \frac{h\nu}{k}[J(T_\mathrm{ex})-J(T_\mathrm{bg})]
    \times[1-\mathrm{exp }(-\tau)]f, \label{eq-Tex}
    \end{eqnarray}
    where $J(T) = [\mathrm{exp}(h\nu/k T)-1]^{-1}$, $T_{\rm bg}=2.73$ K is the temperature of the cosmic background radiation, and $f$ is the beam-filling factor. Under LTE condition, the optical depths of \co\ and \tco\ can be straightforwardly obtained from comparing the measured brightness temperatures:
    \begin{equation}
    \frac{T_r{\rm (^{12}CO)}}{T_r{\rm (^{13}CO)}}\approx\frac{1-{\rm exp} (-\tau_{12})}{1-{\rm exp} (-\tau_{13})}. \label{eq-tau}
    \end{equation}
    Here, $\tau_{12}/\tau_{13}={\rm [^{12}CO]/[^{13}CO]}$. The isotope ratio of ${\rm ^{12}C/^{13}C}$ for each source has been deduced from the following equation \citep{pin13},
    \begin{equation}
      \frac{\mathrm{^{12}C}}{\mathrm{^{13}C}} = 4.7\frac{R_\mathrm{gal}}{\mathrm{kpc}}+25.05, \label{eq-13co-abundance}
    \end{equation}
    where $R_\mathrm{gal}$ is the Galactocentric distance in kpc. With the derived optical depth, the excitation temperature can be reached from \autoref{eq-Tex} based on the intensity of \co.

    Under LTE condition, the column density of a linear molecule can be expressed as
     \begin{eqnarray}
    N&=&\frac{3k}{8\pi^3\nu\mu^2S}\frac{Q_\mathrm{rot}}{g_{J+1}}
    \frac{\mathrm{exp}\left(\frac{E_\mathrm{up}}{k T_\mathrm{ex}}\right)}{J(T_\mathrm{ex})} \nonumber \\
    & &\times\frac{1}{J(T_\mathrm{ex})-J(T_\mathrm{bg})}\frac{\tau}{1-\mathrm{exp}(-\tau)}\int T_\mathrm{r}~dv \label{eq-lte-density}.
    \end{eqnarray}
    Here, $\mu$ is the permanent dipole moment of the molecule. For the $1-0$ transition, the line strength $S = \frac{J+1}{2J+3} = \frac{1}{3}$, the degeneracy $g_\mathrm{J+1} = 2J+3 = 3$, where $J$ is the rotational quantum number of the lower state. We have followed \citet{mcd88} to write the partition function as $Q_\mathrm{rot} = \frac{k T}{hB}+\frac{1}{3}$ where $B$ is the rotational constant. The molecular parameters have been adopted from the Cologne Database for Molecular Spectroscopy (CDMS)\footnotemark[2].
    
     {
    In the calculation, the excitation temperatures of \tco~and \ceo~have 
    been assigned to be equal to that of \co. This would be reasonable while 
    considering the fact that \co, \tco~and \ceo~are well coupled in ECCs 
    \citep{wu12}. The optical depths and column densities of \tco~have been 
    obtained from Equations \ref{eq-tau} and \ref{eq-lte-density}. For 
    \ceo, we derived optical depths and column densities from Equations 
    \ref{eq-Tex} and \ref{eq-lte-density}. 
    }

    \footnotetext[2]{\url{http://www.astro.uni-koeln.de/cdms/}}

    For lacking observations of transitions from an isotopologue and a higher 
    excitation level, the excitation temperatures of \hcop~cannot be reliably 
    derived. Additionally, optically thin feature for most sources 
    (see \autoref{tb-derived}) further prevented us obtaining estimates 
    using \autoref{eq-Tex} under an optically thick assumption. 
    In this work, we assumed the excitation temperatures of 
    \hcopone~to be  {its upper energy in Kelvin 
    ($E_\mathrm{u}/k=4.28$ K)}. Then the optical depths 
    and column densities were derived from Equations \ref{eq-Tex} and 
    \ref{eq-lte-density}.  {We note that the column 
    densities obtained here are lower limits 
    \citep[e.g.,][]{hat98,mie12,mie14} due to the use of upper 
    energy as the excitation temperatures.}

    For HCN, the excitation temperatures were straightforwardly derived from \autoref{eq-Tex} using optical depths obtained from HFS fitting (see section \ref{sec:obs-para} and \autoref{tb-obs-dense}).
    For the derivation of column densities of HCN, \autoref{eq-lte-density} was used by multiplying a factor of 1.8 which comes from the intensity ratios among the main and satellite components.

\begin{figure}
    \centering
    \includegraphics[width=0.45\textwidth]{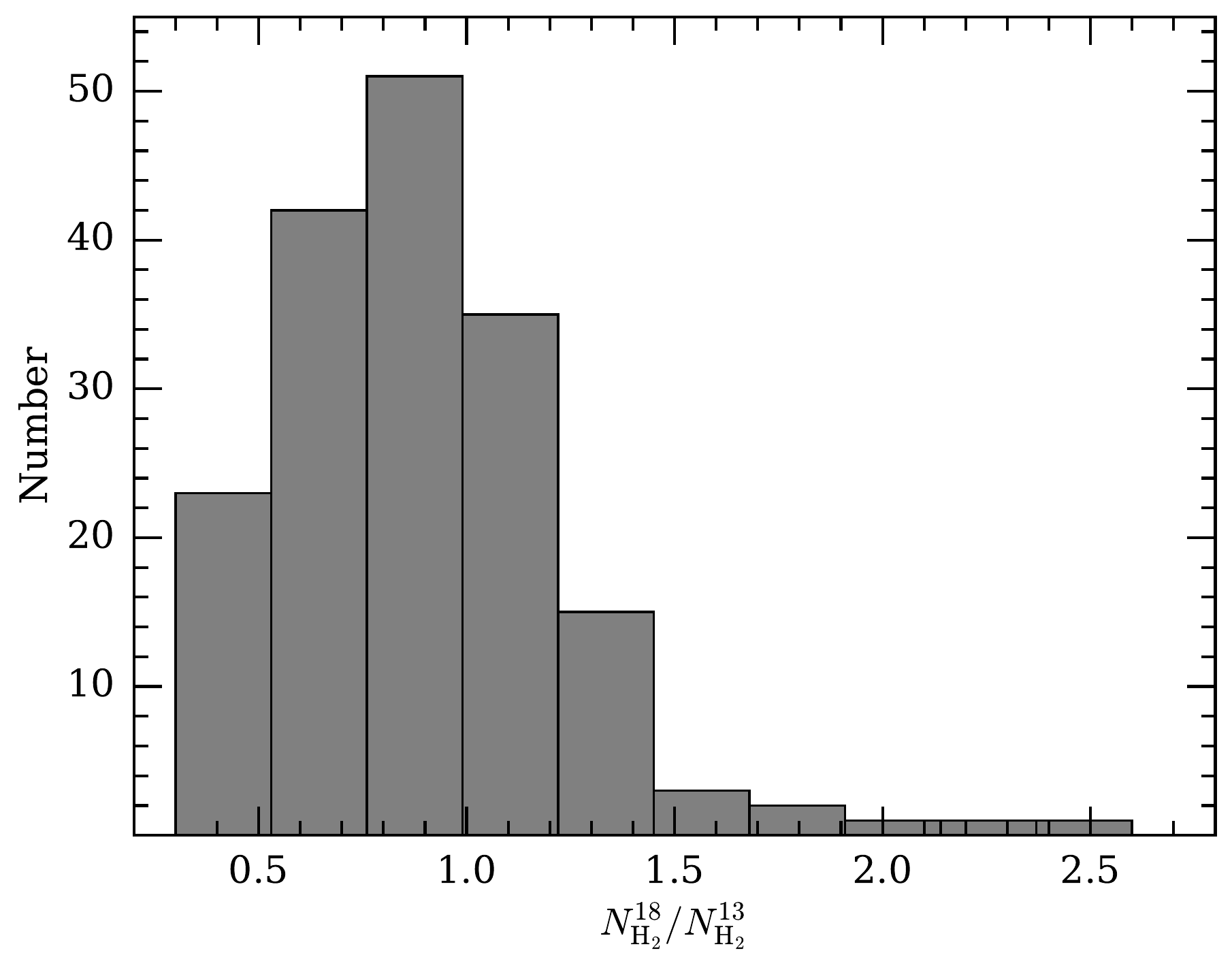}
    \caption{Distribution of the $N_\mathrm{H_2}^{18}/N_\mathrm{H_2}^{13}$ ratios 
    for the 189 sources with \ceo~detection.}\label{fig:dist_col}
    \end{figure}

    The column densities of CO for the 189 cores with detection in \ceo~were calculated from that of \ceo~with a consideration of isotope ratio of $\mathrm{^{16}O/^{18}O}$ as a function of Galactocentric distance \citep{wil94},
     \begin{equation}
       \frac{\mathrm{^{16}O}}{\mathrm{^{18}O}}=58.8~\frac{R_\mathrm{gal}}{\rm kpc}+37.1.
     \end{equation}
     For sources without detection in \ceoone, the CO column densities were obtained from that of \tco~by adopting isotope ratio given in \autoref{eq-13co-abundance}. Then the H$_2$ column densities were derived with the consideration of CO abundances as a function of  Galactocentric distances \citep{fon12},
    \begin{equation}
      X_\mathrm{CO} = 8.5\times10^{-5}~\mathrm{exp}\left(1.105 - 0.13~\frac{R_\mathrm{gal}}{\rm kpc}\right). \label{eq-co-abundance}
    \end{equation}
     {In order to test the consistency between the two methods, we have obtained 
    two sets of H$_2$ column densities of the 189 sources from \ceo~and \tco~and  denoted them 
    as $N_\mathrm{H_2}^{18}$ and $N_\mathrm{H_2}^{13}$. As shown in \autoref{fig:dist_col}, 
    the $N_\mathrm{H_2}^{18}/N_\mathrm{H_2}^{13}$ ratios
    are ranging from 0.3 to 2.6 with a mean of 0.9 and a standard derivation of 0.3, indicating that the 
    H$_2$ column densities derived from \tco~and \ceo~are consistent to each other.}

     {
    One should note that the derived H$_2$ column densities would suffer from large uncertainties due to the 
    abundance variations of CO. As noted in the literature the Galactic gradients of abundances 
    (Equations 4, 6, and 7) would be valid in the Galactocentric distances between 3 kpc and 10 kpc 
    where CO has significant emission \citep{wil94,pin13}. As shown in \autoref{fig:mw-faceon}, most sources
    reside inside 10 kpc from the Galactic center. For these targets inside the 10 kpc circle, abundance 
    ratios of $\mathrm{[^{12}C]/[^{13}C]}$, $\mathrm{[^{16}O]/[^{18}O]}$ and $\mathrm{[CO]/[H_2]}$ have been
    obtained from Equations 4, 6, and 7. The $\mathrm{[^{12}C]/[^{13}C]}$ ranges from 39 
    to 72, $\mathrm{[^{16}O]/[^{18}O]}$ from 213 to 625, and $\mathrm{[CO]/[H_2]}$ from $0.7\times10^{-4}$
    to $1.7\times10^{-4}$. For sources with Galactocentric distances larger than 10 kpc, the abundance ratios
    at $R_\mathrm{gal}=10$ kpc have been adopted. We compared the $\mathrm{[^{12}C]/[^{13}C]}$ obtained from 
    \autoref{eq-13co-abundance} ratios to that from \citet{wil94} and found that the differences are smaller than 25\%. We estimate
    that the variations of abundance ratios would introduce uncertainties smaller than a factor of one for
    sources with $R_\mathrm{gal}<10$ kpc. The uncertainties for sources with $R_\mathrm{gal}>10$ kpc would 
    be larger. 
    }

    The resulting optical depths of \tco, \ceo~and \hcop, excitation temperatures of CO, column densities of \tco, \ceo, \hcop, and HCN are given in columns 4--11 of \autoref{tb-derived}. The statistics of some derived parameters are presented in \autoref{tb-derived-stat}. The excitation temperatures in the CO-selected cores with \hcop and/or HCN detection range from 7.6 to 34.4 K, significantly higher than the values in \citet{wu12} and the dust temperatures \cite[7--19 K, ][]{pla11d}. The differences may be due to the sources in this work are relatively dense regions in clumps studied by \citet{pla11d} and \citet{wu12}. The mean excitation temperature is about 12.54 K approximately equal to the mean dust temperature \citep[13 K,][]{pla11d} and slightly higher than the value in \citet{wu12}. The excitation temperatures of these source with \hcop~and/or HCN detection are significantly higher than that of CO-selected cores in the Taurus, Perseus, and California complexes \citep{men13}, but similar to that of CO-selected cores in the Orion complex \citep{liu12}. The column densities of H$_2$ cover the range of $(1.5-70.9)\times10^{21}$ cm$^{-2}$ with a mean value of $1.3\times10^{22}$ cm$^{-2}$, much higher than those of CO-selected cores in the Orion, Taurus, Perseus, and California complexes \citep{liu12,men13}. This is a natural result because the sources with detection in \hcop~and/or HCN are relative denser ones.

    \section{Discussions}\label{s-discussions}

        \begin{figure*}
    \centering
    \includegraphics[width=0.9\textwidth]{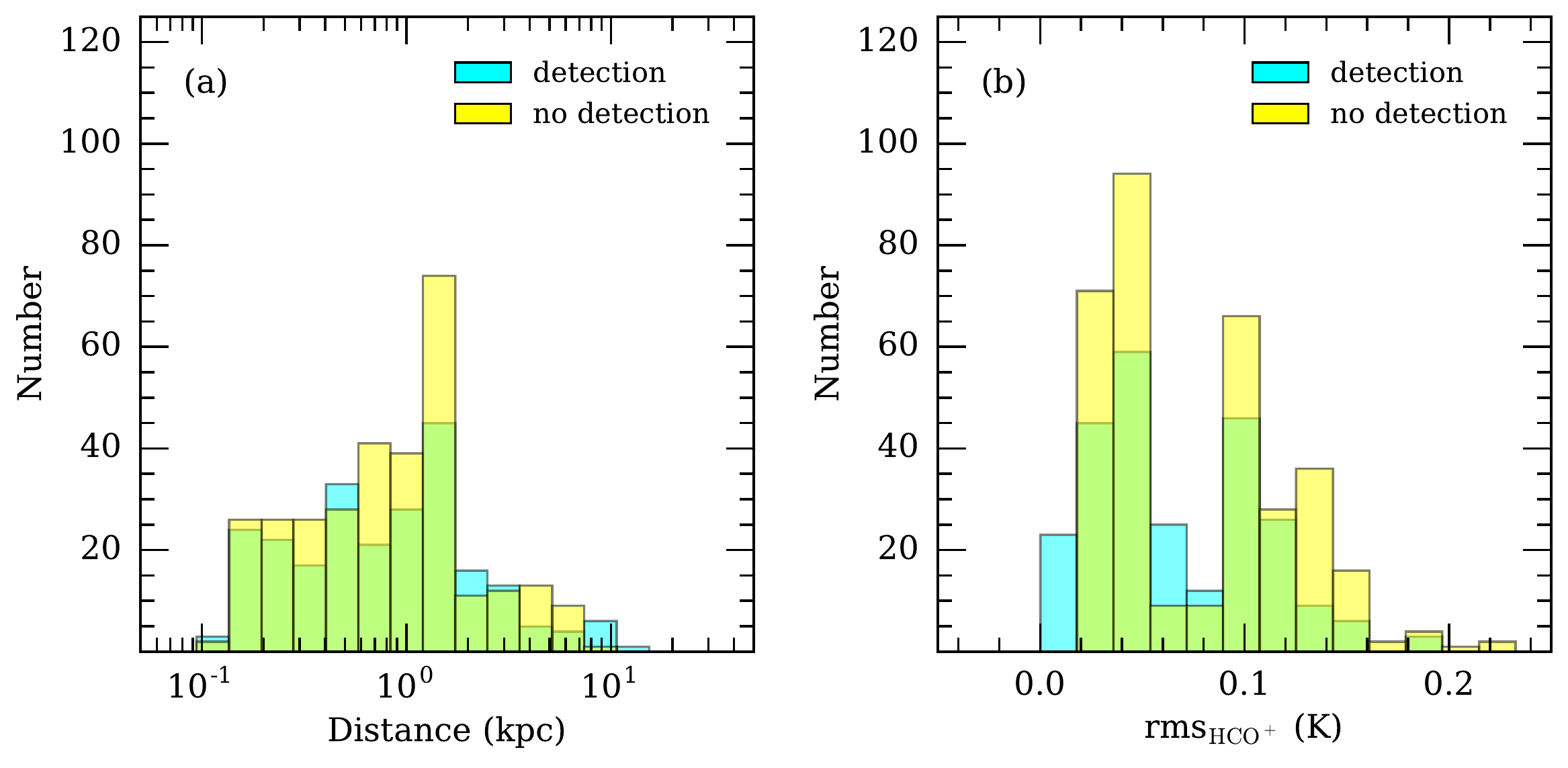}
    \caption{Distributions of distances ($a$) and RMS noise of \hcop~($b$) for sources with 
    detection (cyan) in ether \hcop~or HCN, and sources without detection (yellow) in neither
    \hcop~nor HCN. The green means an overlay between the two groups.}\label{fig:comp-dist-sigma}
    \end{figure*}

     {
    In order to assure the detection of \hcop~and/or~HCN is not purely distance and 
    sensitivity limited, we have plotted histograms of distances and RMS noises of 
    \hcop~for CO-selected cores with and without detection.  As shown in Figure 
    \ref{fig:comp-dist-sigma}(a), the distances for sources with and without 
    detection have a similar distribution. The sources with detection in dense gas 
    tracers have distances ranging from 0.1 to 14.7 kpc with a mean of 1.4 kpc 
    and a median of 0.8 kpc. For sources without detection, the values range from 
    0.1 to 10.2 kpc with a mean of 1.2 kpc and a median value of 0.9 kpc. The 
    similar mean and median values for source with and without 
    detection suggest that distances
    have not induced biases to the detection of 
    dense gas tracers in the CO-selected cores. 
    As shown in Figure \ref{fig:comp-dist-sigma}(b), the distributions of RMS 
    noises of sources with and without detection are similar with an excess for 
    non-detection sources at the high noise end ($\mathrm{rms}>0.12$ K). The 
    fractions of high noises are 0.18 and 0.07 for sources with and without 
    detection. Nevertheless, more than 80\% CO-selected cores without detection 
    have RMS noises as low as the ones with detection. To further check the influence of
    the large range of sensitivity on the detection, we have separated the observed sources
    into high- and low-sensitivity groups using the median RMS noise
    ($\sim0.06$ K) as a threshold. The detection rates for the high- and low-sensitivity
    groups are 38.7\% and 40.5\%, respectively. Such consistency, together with
    the similarity between distributions of RMS noises of sources with and without
    detection, suggests that the detection of \hcop~and/or~HCN is not sensitivity limited.  
    }

    \subsection{Kinematics}

    \begin{figure}[tbp]
    \centering
    \includegraphics[width=0.45\textwidth]{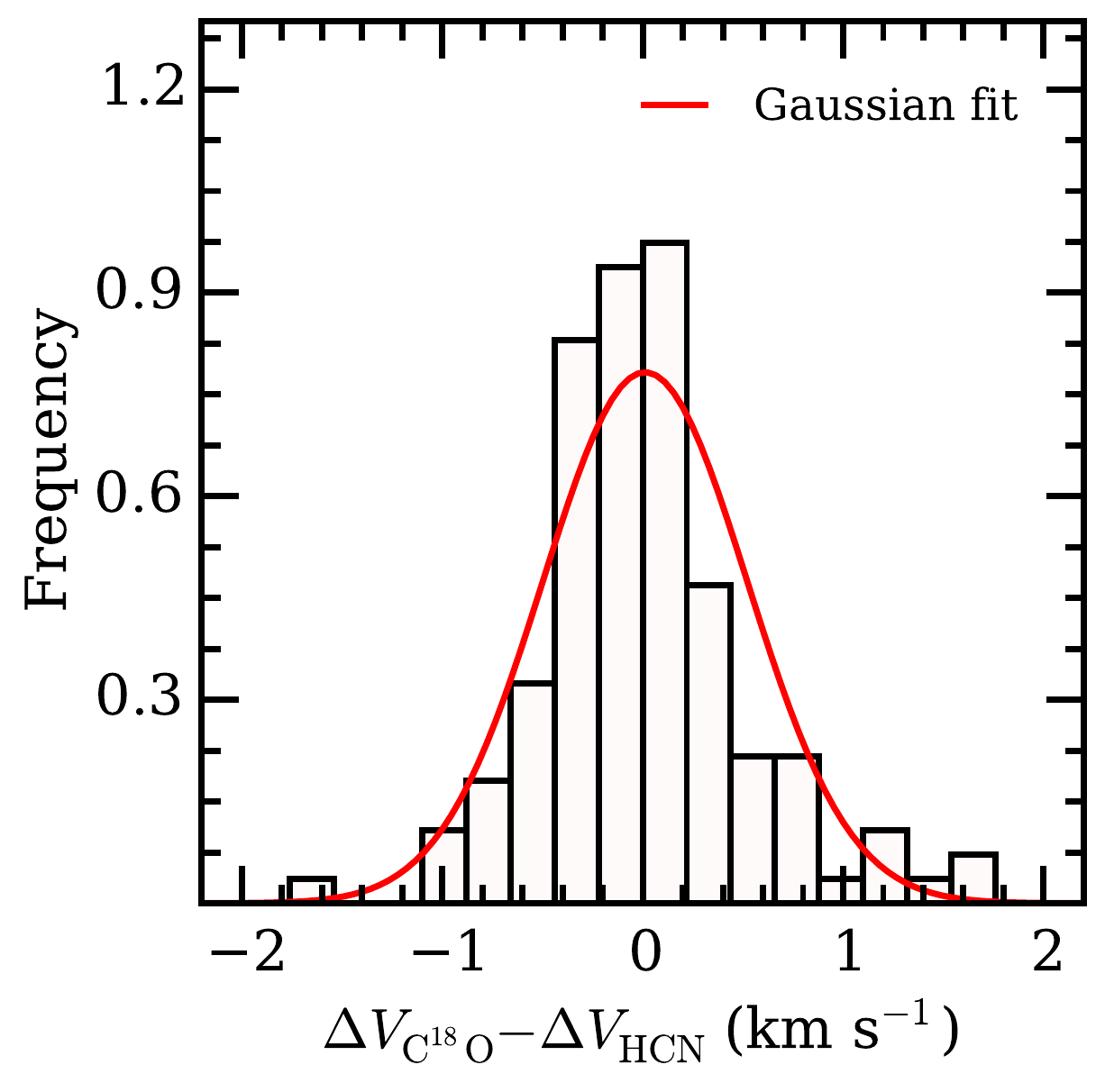}
    \caption{Distributions between differences of line widths of \ceo~and HCN.}\label{fig:widthDif-distri}
    \end{figure}

    We have carefully inspected the spectra of sources with detection in \hcop~and/or 
    HCN to find that all targets show single-peaked profiles except for seven ones 
    showing self-absorption features (see \autoref{fig:spectra}). Among the seven 
    exceptions, two are red profiles (i.e. lines with higher peaks skewed to the red 
    sides; G084.79-01.11A1 and G120.16+03.09A1) and four blue profiles (G114.56+14.72A1, 
    G126.49-01.30A2, G172.85+02.27A1, and G226.36-0-0.50A1). The other one shows 
    symmetric double-peaked profile with a dip at the line center (G120.98+02.66A1). 
    Asymmetric double-peaked profiles are always regarded as effective tracers of 
    kinematics in star-forming cores with blue profiles linked to inward motions 
    \citep[e.g., collapse or infall,][]{zho93,wu03,eva05,wu05} and red profiles 
    for outward motions \citep[e.g., outflow or expansion][]{wu07,yua13}. The 
    detection rates of blue and red profiles are 1.2\% and 0.8\%, significantly 
    smaller than the values in surveys toward low- and high-mass star-forming 
    regions. Based on results in the literature, \citet{eva03} found the detection 
    rates of blue (red) profiles in Class -I, 0, and I low-mass star-forming 
    regions are 35\% (6\%), 33\% (5\%),  and 50\% (19\%), respectively. Among the 
    48 high-mass clumps mapped in HCN $J=3-2$ and an optically thin line 
    (H$^{13}$CN $J=3-2$ or C$^{34}$S $J=5-4$), \citet{wu10} reported detection 
    rates of 44\% and 31\% for blue and red profiles.  {We note that
    the difference between detection rates of asymmetric profiles in this work and
    that in the literature are not due to a difference in S/N of the observations, because
    the detection rates are still significantly low ($<3\%$) even if we only consider
    sources with $S/N>10$.}
    The small detection rates of asymmetric double-peaked profiles 
    in this work indicate rareness of bulk-motions in these dense 
    CO-selected cores, suggesting they are relatively quiescent compared 
    to classical star-forming regions.

    Another indicator of kinematics that can be extracted from single-point observations could be the line widths. The mean widths of \co, \tco, and \ceo~of the CO-selected cores with valid detection in \hcop~and/or HCN are 2.95, 1.67, and 1.08 \kms, relatively larger than that in \citet{wu12} in which they are 2.0, 1.3, and 0.8 \kms. The spectra are also slightly wider than those of CO-selected cores in Orion (0.9 \kms~for \tco), Taurus (1.8 \kms~for \co~and 1.1 \kms~for \tco), and California (2.2. \kms~for \co~and 1.4 \kms~for \tco) molecular complexes \citep{liu12, men13}. Slightly larger line widths detected in this work may be attributed to that the cores with detection in \hcop~and/or HCN are significantly denser than those in Taurus, California and Orion complexes (see section \ref{sec:derived-para}) and higher ram pressures make additional contribution to the line widths. Another possibility could be that the CO-selected cores with detection in \hcop~and/or HCN are more turbulent.

    Using data of multiple lines from distinct species with significantly different critical densities, we can statistically examine the difference of kinematics in separate regions. Here, we mainly compare the line widths of \ceo~and HCN which, in general, are optically thin.  {The typical source-averaged volume 
    density is $4.6\times10^3$ cm$^{-3}$ 
    (see section \ref{s-obs}), indicating that there may be a significant volume with density 
    above $10^4$ cm$^{-3}$ in these sources.} With a critical density no larger than $10^4$ cm$^{-3}$ , \ceoone~mainly originates from the exterior of a core. Meanwhile, \hcnone~is a good tracer of the inner region due to its significantly higher critical density of~$>10^5$ cm$^{-3}$. The line widths of \ceo~are in the range of 0.17--3.39 \kms~with a mean value of 1.08 \kms, while those of HCN ranging from 0.69 to 2.70 \kms~with a mean of 1.06.  {The typical optical depths are 0.2 and 0.3 for 
    \ceoone~and \hcnone, respectively. Such small optical depths can only broaden the line widths 
    by a factor of  
    $<10\%$ which has been estimated using Equation 3 of \citet{phi79}. } We note that the minimum value for HCN is only an upper limit due to the poor spectral resolution of our observations (see \autoref{s-obs}). The mean widths of \ceo~and HCN are almost equal to each other.  The distribution of difference between line widths of \ceo~and HCN, which can be fitted using a Gaussian with a $\mu=0.01$ \kms~and a $\sigma=0.51$ \kms, is shown in \autoref{fig:widthDif-distri}. Intriguingly, the width difference is very small with statistical significance. This is contrary to the results in \citet{wu10} in which larger widths of transitions with higher critical densities are treated as an indication of larger turbulence in the inner region of dense cores. The almost equivalence between widths of \ceo~and HCN suggests that the inner regions of CO-selected cores are not more turbulent than the exterior, different from the situation in high-mass star-forming cores where the interior is of larger turbulence \citep[e.g.,][]{wu10,gar15}.  {Compared to the two optically thin lines (i.e., \ceoone~and \hcnone), the lines of \tcoone~
    are relatively 
    broader with widths ranging from 0.59 to 6.25 \kms~with a mean value of 1.67 . The typical difference between
    line widths of \tco~and HCN is 0.65 \kms, and the value is 0.42 \kms~even accounting for the broadening attributed
    to optical depth. This indicates the surrounding clouds are more turbulent.}

    \subsection{Coupling of Different Species}

    \begin{figure}
    \centering
    \includegraphics[width=0.45\textwidth]{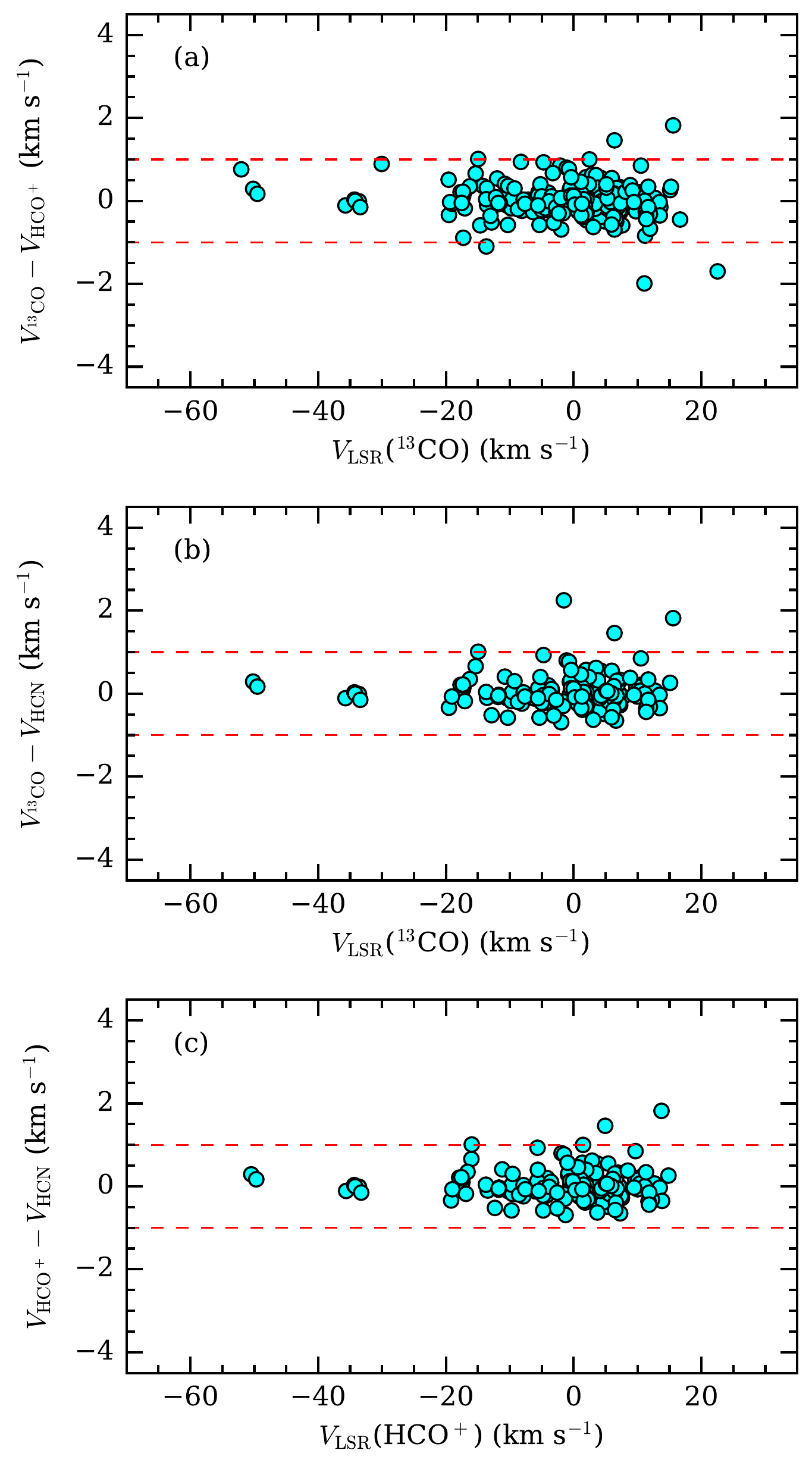}
    \caption{Correlations among velocity centroids of different species. \emph{(a)} $V_\mathrm{LSR}$ (\hcop) vs. $V_\mathrm{LSR}$ (\tco); \emph{(b)} $V_\mathrm{LSR}$ (HCN) vs. $V_\mathrm{LSR}$ (\tco); \emph{(c)} $V_\mathrm{LSR}$ (HCN) vs. $V_\mathrm{LSR}$ (\hcop). The lower panels shows the velocity differences. }\label{fig:vel-relation}
    \end{figure}

    \begin{figure*}
    \centering
    \includegraphics[width=0.9\textwidth]{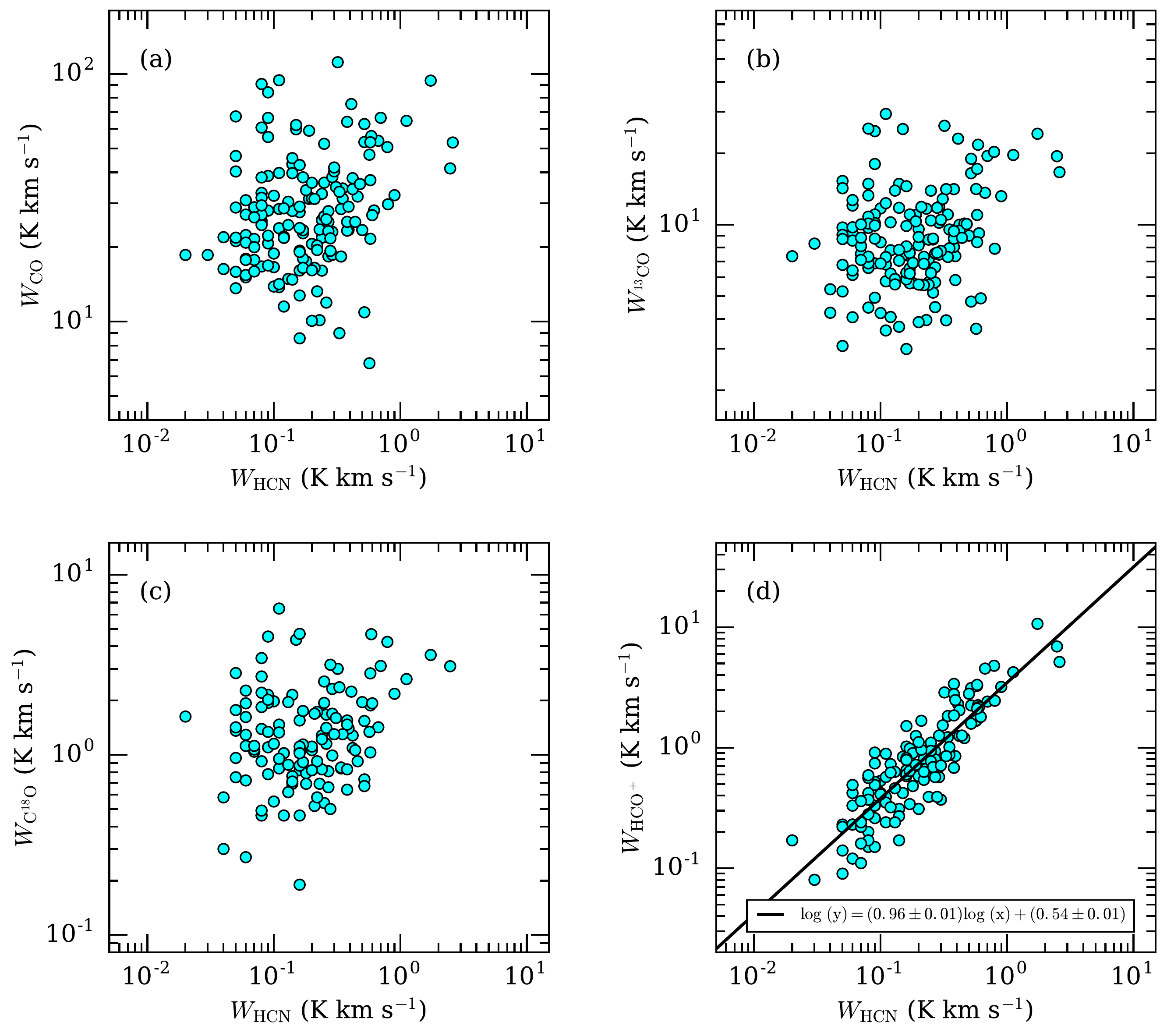}
    \caption{Velocity integrated intensities of CO \emph{(a)}, \tco~\emph{(b)}, \ceo~\emph{(c)}, and \hcop~\emph{(d)} as functions of that of HCN.}\label{fig:inten-relation}
    \end{figure*}

    In the survey toward 674 \emph{Planck} cold clumps, \citet{wu12} found that the central velocities of transitions from CO and its isotopologues are strikingly consistent with each other. This feature could be reasonable in quiescent regions, for CO and its isotoplogues should be well coupled with each other without active disturbance. In this work correlations among central velocities of \tco, \hcop~and HCN, which are shown in \autoref{fig:vel-relation}, have been examined. The differences of central velocities between \tco~and \hcop~($V_\mathrm{^{13}CO}-V_\mathrm{HCO^+}$), \tco~and HCN ($V_\mathrm{^{13}CO}-V_\mathrm{HCN}$), \hcop~and HCN ($V_\mathrm{HCO^+}-V_\mathrm{HCN}$) have mean values of $0.006\pm0.4$, $0.05\pm0.4$, $0.002\pm0.6$ \kms. Intriguingly, the differences are strikingly small, which may be an indication of quiescent features for these CO-selected cores. Noticeably, the deviations of ($V_\mathrm{^{13}CO}-V_\mathrm{HCO^+}$), ($V_\mathrm{^{13}CO}-V_\mathrm{HCN}$), and ($V_\mathrm{HCO^+}-V_\mathrm{HCN}$) from zero seems to be consistent with the critical densities of the $J=1-0$ transitions of \tco~($\sim10^3$ cm$^{-3}$), \hcop~($\sim7\times10^4$ cm$^{-3}$), and HCN~($\sim5\times10^5$ cm$^{-3}$). If HCN, \hcop, and \tco~trace gases from the interior to the exterior, the deviation of velocity difference from zero can also reflect the spatial correlation among different species. For instance, the small deviation of ($V_\mathrm{HCO^+}-V_\mathrm{HCN}$) may indicate that \hcop~is better coupled with HCN than CO.

    Shown in \autoref{fig:inten-relation} are relations of velocity integrated intensities of \co, \tco, \ceo, and \hcop~with that of HCN. The intensities of CO and its isotopologues seem not correlated with that of HCN. However, the relationship between \hcop~and HCN follows a power law of $W_\mathrm{HCO^+}=2.3\times W_\mathrm{HCN}^{0.8}$ with  {a coefficient of determination of $R^2=0.7$ indicating about 70\% of the data points can be well represented by the fitted linear relation in the log-log space}. Such prominent correlation gives additional evidence to the coupling between \hcop~and HCN. Strong correlation between \hcop~and HCN was also detected in \citet{liu13b} and interpreted as an indicator of a tight chemical connection.

    \subsection{Abundances of \hcop~and HCN}

    The statistics of column densities and fractional abundances of \hcop and 
    HCN are given in \autoref{tb-derived-stat}. The column densities of \hcop~are 
    in the range of $(0.07-24.16)\times10^{12}$ cm$^{-2}$ with a mean value of 
    $1.69\times10^{12}$ cm$^{-2}$ , while that of HCN ranging from $0.24\times10^{12}$ cm$^{-2}$ to $16.15\times10^{12}$ cm$^{-2}$ with a mean of $1.57\times10^{12}$ cm$^{-2}$. The abundances of \hcop~(HCN) range from $3.0\times10^{-12}$ ($1.3\times10^{-11}$) to $1.5\times10^{-9}$ ($5.6\times10^{-10}$) with a mean value of $1.5\times10^{-10}$ ($1.4\times10^{-10}$) which are significantly lower than values in low- to high-mass star-forming regions where the abundances of both \hcop~and HCN are in the range of $10^{-10}-10^{-8}$ \citep{dut97,hir98,vas11,san12,liu13b,ger14,mie14}. The low abundances in this work could be attributed to poor spatial resolutions of our observations which have led to severe effect of beam dilution. The CO-selected cores with valid detection in \hcop~and/or HCN have distances ranging from 0.1 pc to 14.7 kpc with a mean of 1.4 kpc. The corresponding physical scales covered by one beam range from 0.03 pc to 4.3 pc with a mean of 0.35 pc. However the region where the gas is dense enough for \hcop~and HCN to be detectable would be smaller than 0.1 pc. Thus the measured column densities of \hcop~and HCN have been severely diluted by a factor of tens on average, even by a factor of $10^3$ for the farthest sources.  {We note that 
    the H$_2$ column densities, which have been used to obtain fractional abundances of \hcop~and HCN, were
    estimated based on CO observations. As aforementioned, the variation of $\mathrm{[^{12}C]/[^{13}C]}$, 
    $\mathrm{[^{16}O]/[^{18}O]}$ and $\mathrm{[CO]/[H_2]}$ ratios would lead to large uncertainties to the 
    results. Further dust observations with sufficient resolution are needed to better constrain the abundances.
    }

    \begin{figure*}
    \centering
    \includegraphics[width=0.8\textwidth]{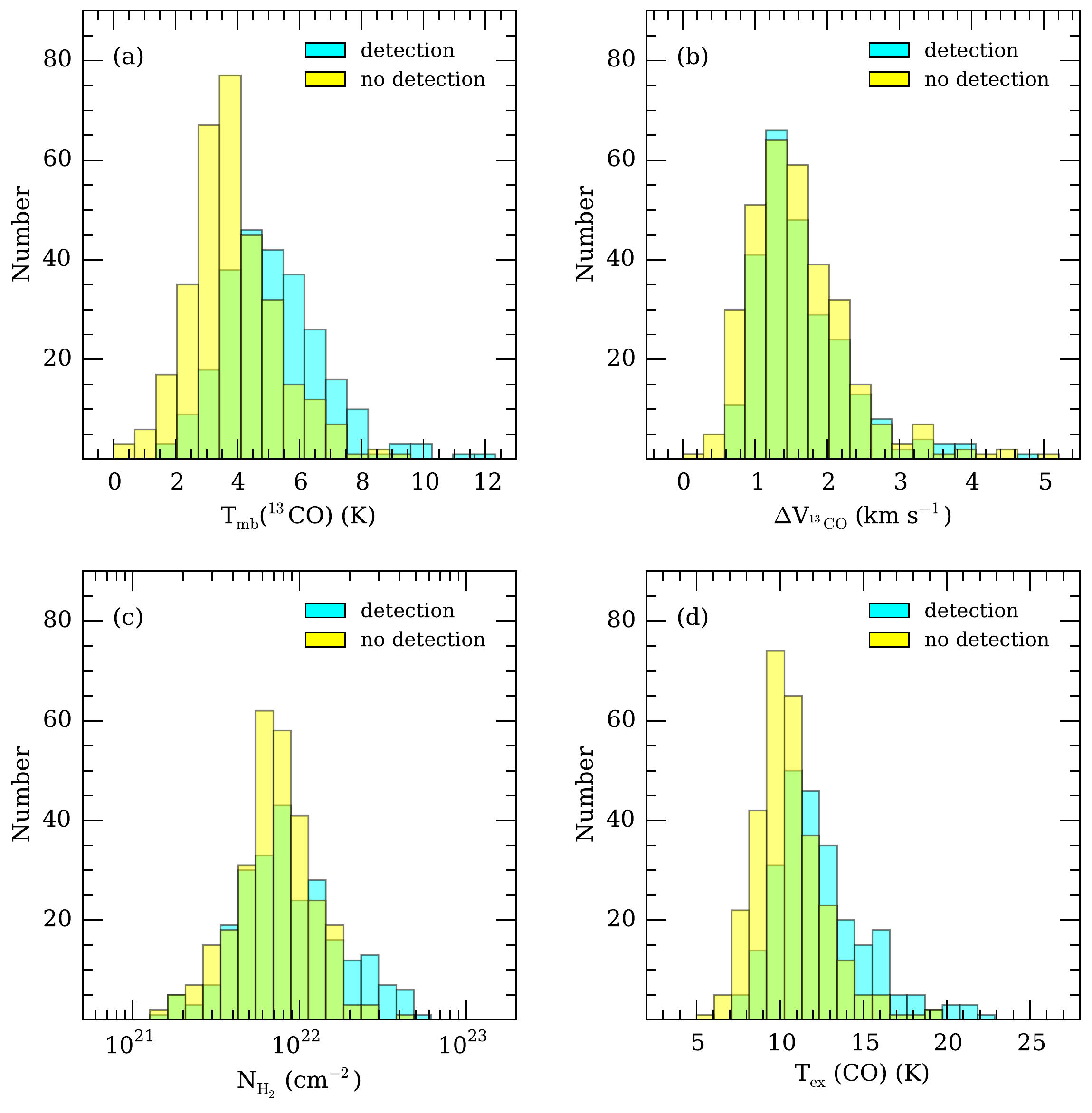}
    \caption{Distributions of velocity-integrated intensities of \tco~($a$), line widths of \tco~($b$), H$_2$ column densities and excitation temperatures of CO for sources with 
    detection (cyan) in ether \hcop~or HCN, and sources without detection (yellow) in neither
    \hcop~nor HCN. The green means an overlay between the two groups.}\label{fig:para-comp-co}
    \end{figure*}

    \begin{figure*}
    \centering
    \includegraphics[width=0.8\textwidth]{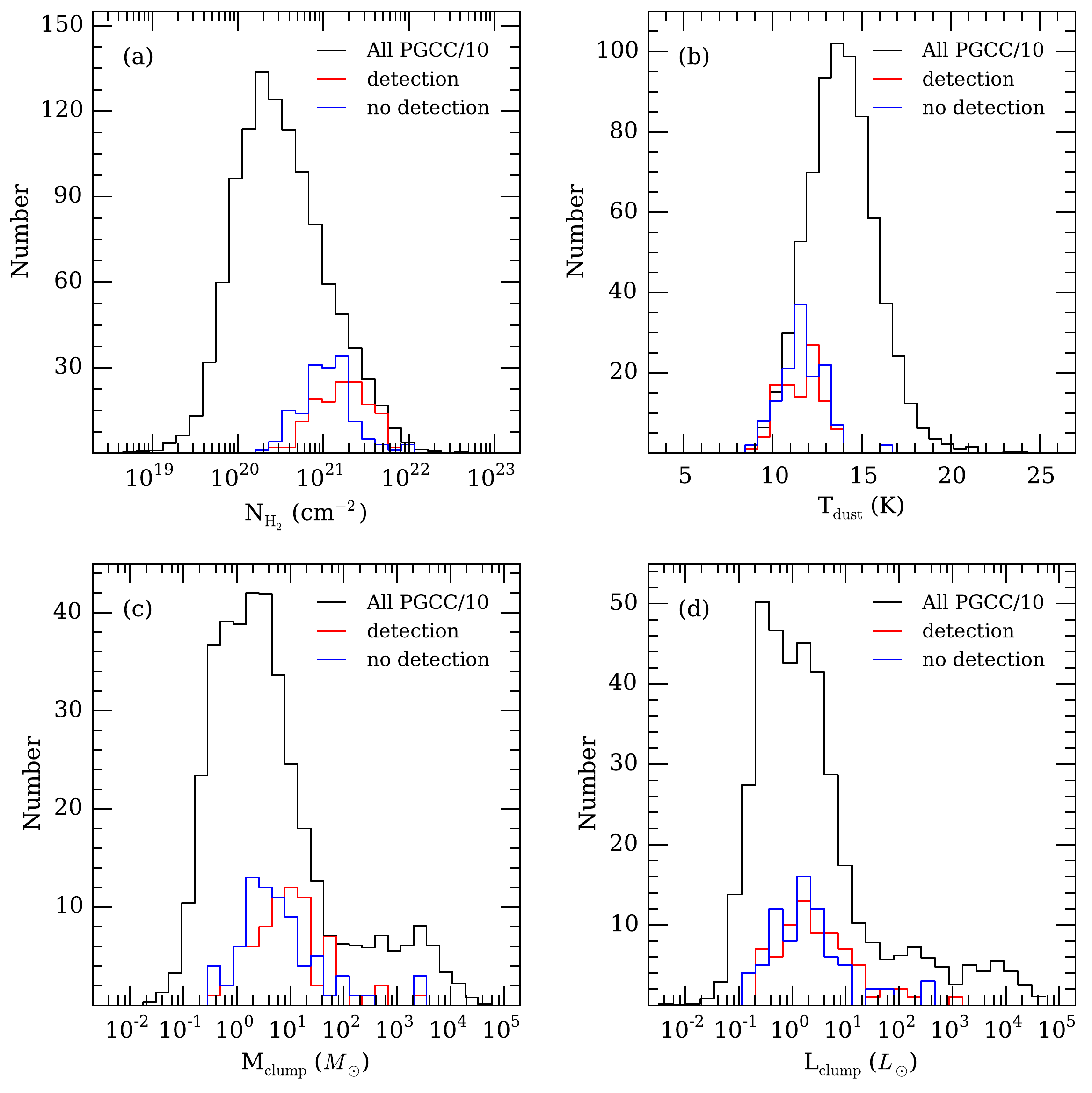}
    \caption{Distributions of H$_2$ column densities, dust temperatures, masses and luminosities of all PGCCs (black) and the ones consisting CO-selected cores with (red) and without (blue) detection in \hcop~and/or HCN.  {We note that
        all the parameters are from the PGCC catalog \citep{pla15}}}\label{fig:par-com-pgcc}
    \end{figure*}
    
    \subsection{PGCCs with Dense Gas}\label{sec:pgccs}
    
    As the first survey of dense gas towards $Planck$ cold clumps, the statistical results in this work may lead us to determine whether there are any hints, in the many physical parameters given in the PGCC catalog, that can point to the presence of dense gas. 
    
    For sources detected in dense gas tracers, what makes them stand out of the rest? To address this question, we have examined some parameters from CO observations i.e., peak intensities of \tco, line widths of \tco, H$_2$ column densities, and excitation temperatures of CO. Shown in \autoref{fig:para-comp-co} are histograms of these inspected parameters. The CO-selected cores with and without detection in \hcop~and/or HCN share common distributions for line widths of \tco. The peak intensities of \tco~and excitation temperatures of CO for sources detected in \hcop~and/or HCN are systematically higher than that of sources without detection. For H$_2$ column densities, the two groups of sources share a similar distribution at low- to intermediate densities. There is a prominent excess at H$_2$ column densities higher than $2\times10^{22}$ cm$^{-2}$ for sources detected in dense gas tracers. In general, sources detected in \hcop~and/or HCN would have higher CO excitation temperatures and, probably, higher column densities compared to the ones without detection.
    
    In order to pinpoint which parameter(s) given in the PGCC catalog can point to the 
    presence of dense gas, we have revisited H$_2$ column densities, dust temperatures, 
    masses and luminosities of all PGCCs (black) and the ones consisting CO-selected cores 
    with (red) and without (blue) detection in \hcop~and/or HCN. The histograms and statistics of 
    these parameters are presented in \autoref{fig:par-com-pgcc} ahd \autoref{tb-pgcc-par}. In general, 
    the PGCCs containing CO-selected cores (for both with and without detection in dense gas tracers) 
    are denser and colder compared to the whole PGCC sample. There is no surprise for this point
    because the CO-selected
    cores were identified from ECCs which constitute the high-S/N and cold tail of PGCC objects. 
    PGCCs containing CO-selected cores detected in \hcop~and/or HCN are slightly denser compared to
    the ones with no detection. There is no significant difference between distributions of dust temperatures for these 
    two subsamples. As shown in \autoref{fig:par-com-pgcc}(c), the clump masses of 
    the full PGCC sample and the PGCCs without detection in dense gas have similar distributions, 
    while that of the the PGCCs with detection in \hcop~and/or HCN slightly deviating to the high mass end
    with larger median values. For luminosities, the full PGCC sample and the other two subsamples share 
    similar distributions and the median values are also consistent in a factor of one 
    (see \autoref{fig:par-com-pgcc} and \autoref{tb-pgcc-par}). Although there are no significant differences for
    parameters among distinct samples, we still tentatively found that 
    PGCC sources containing dense gas would be denser and more massive compared to the remaining pool. 
     {On the basis of the statistics presented in \autoref{tb-pgcc-par}, we tentatively estimate that 
    there may be about 1000 PGCC objects having sufficient reservoir of dense gas
    to form stars. }

    \section{Conclusions}\label{s-conclusions}

    We have carried out single-pointing observations in the $J=1-0$ transitions of \hcop~and HCN towards 621 CO-selected molecular cores associated with \emph{Planck} cold clumps. Among them, 250 sources have valid detection in either \hcop~and/or HCN, including 138 cores detected in both lines, 92  cores only in \hcop, and 20 cores only in HCN. Spectra of the $J=1-0$ transitions of \co, \tco, and \ceo~for the 250 cores were extracted from previous mapping observations to construct a multi-line data set to reveal some kinematic and chemical properties.

    Spectra of \hcopone~and \hcnone~of all sources show single-peaked profiles except for seven ones showing self-absorption features. Among the seven exceptions, two are red profiles, four blue profiles. The other one shows a symmetric double-peaked profile with a dip at the line center. The low detection rate of asymmetric profiles suggests the CO-selected cores are more quiescent compared to classical star-forming regions.

    The difference between line widths of \ceo~and HCN are close to zero with a small standard deviation, indicating that the inner regions of CO-selected cores are not more turbulent than the exterior. This is different from the situation in high-mass star-forming cores where the interior is of larger turbulence \citep{wu10,gar15}.

    The central velocities of \hcop~are more consistent with that of HCN than \tco. The velocity-integrated intensities of \hcop~are well correlated with that of HCN. These features indicate that gaseous \hcop~is better coupled with HCN than CO and its isotopologues.

    The H$_2$ column densities of sources with detection in \hcop~or~HCN~have been deduced from spectra of \ceo~or \tco, and are in the range of $(1.5-70.9)\times10^{21}$ cm$^{-2}$ with a mean value of $1.3\times10^{22}$ cm$^{-2}$, significantly higher than those of CO-selected cores in Orion, Taurus, Perseus, and California complexes.

    The estimated fractional abundances of \hcop~(HCN) range from $3.0\times10^{-12}$ ($1.3\times10^{-11}$) to $1.5\times10^{-9}$ ($5.6\times10^{-10}$) with a mean value of $1.5\times10^{-10}$ ($1.4\times10^{-10}$) which are significantly lower than values in low- to high-mass star-forming regions. The low abundances detected in this work may be attributed to beam dilution.

    The inspection of some parameters from CO observations suggests that CO-selected 
    cores detected in \hcop~and/or HCN would have higher CO excitation temperatures
    and, probably, higher column densities compared to the ones without detection. 
    Among the parameters given in the PGCC catalog, we found that the H$_2$ column
    density would serve as an indicator of the presence of dense gas in PGCCs. 
    On the basis of the inspection of 
    the parameters given in the PGCC catalog, we suggest that there are about 
    1 000 PGCC objects may have sufficient reservoir of dense gas to form stars.

\begin{acknowledgements}

    This work is supported by the National Natural Science 
    Foundation of China through grants of 11503035, 11573036, 
    11373009, 11433008 and 11403040, 
    the China Ministry of Science and Technology
    under State Key Development Program for Basic Research
    through the grant of 2012CB821800, the International 
    S\&T Cooperation Program of China through the grand of 
    2010DFA02710,  the Beijing Natural Science Foundation 
    through the grant of 1144015. and the Young Researcher 
    Grant of National Astronomical 
    Observatories, Chinese Academy of Sciences. 
    KW acknowledges support from the ESO fellowship and 
    grant WA3628-1/1 through the DFG priority programme 
    1573 `Physics of the Interstellar Medium' 
    (http://www.ism-spp.de). We give our thanks to the staff 
    at the Qinghai Station of Purple Mountain Observatory for 
    their hospitable assistance during the observations and 
    the Key Laboratory for Radio Astronomy of Chinese Academy 
    of Sciences for partial support in the operation of the telescope.

\end{acknowledgements}

\clearpage
  \LongTables
\begin{landscape}

\begin{deluxetable}{lrrrrrrrrrrrrrr}
        \tablecaption{Observational Parameters of \hcop~ and HCN \label{tb-obs-dense}}
    \tablewidth{0pt}
    \tablehead{
    \colhead{Designation} &\colhead{R.A.}& \colhead{Dec.}& \multicolumn{5}{c}{HCO$^+$ $J=1-0$} & & \multicolumn{6}{c}{HCN $J=1-0$}  \\
    \cline{4-8}\cline{10-15} \\
    &&& \colhead{rms} & \colhead{$T_\mathrm{mb}$}& \colhead{$\int T_\mathrm{mb}dv$}& \colhead{$v_\mathrm{lsr}$}& \colhead{FWHM}& &\colhead{rms} & \colhead{$T_\mathrm{mb}$}& \colhead{$\int T_\mathrm{mb}dv$}& \colhead{$v_\mathrm{lsr}$}& \colhead{FWHM} & \colhead{$\tau$}  \\
     & && \colhead{(K)}& \colhead{(K)}&\colhead{(K \kms)} & \colhead{(\kms)}& \colhead{(\kms)} & &  \colhead{(K)}& \colhead{(K)}&\colhead{(K \kms)} & \colhead{(\kms)}& \colhead{(\kms)} &
    }
    \startdata
     G026.45+08.02a1 &18h11m20.92s &  -2d00m22.7s &     0.04 &   0.60(0.07)  &    0.39(0.03) &   6.17(0.02) &  0.60(0.05) &  &  0.03 &   0.14(0.20) &  0.11(0.17) &   6.31(0.10) &  1.10(0.19) &   0.14(0.15) \\  
 G031.26-05.37a1 &19h07m39.03s &  -4d00m27.3s &     0.12 &   0.40(0.18)  &    0.27(0.09) &   0.74(0.12) &  0.64(0.20) &  &  \nodata &   \nodata &  \nodata &   \nodata &  \nodata &   \nodata \\  
 G038.36-00.95a1 &19h04m44.03s &   4d22m51.4s &     0.04 &   0.16(0.04)  &    0.34(0.05) &  17.11(0.16) &  1.98(0.36) &  &  \nodata &   \nodata &  \nodata &   \nodata &  \nodata &   \nodata \\  
 G043.02+08.36a1 &18h39m21.54s &  12d41m35.3s &     0.19 &   1.00(4.20)  &    0.44(0.12) &   4.07(0.05) &  0.41(1.72) &  &  \nodata &   \nodata &  \nodata &   \nodata &  \nodata &   \nodata \\  
 G057.17+03.41a1 &19h24m55.57s &  23d04m21.7s &     0.03 &   0.13(0.13)  &    0.07(0.02) &  12.03(0.12) &  0.54(0.50) &  &  \nodata &   \nodata &  \nodata &   \nodata &  \nodata &   \nodata \\  
 G058.97-01.66b1 &19h47m42.84s &  22d06m07.8s &     0.04 &   0.22(0.39)  &    0.10(0.03) &  24.23(0.07) &  0.41(0.71) &  &  \nodata &   \nodata &  \nodata &   \nodata &  \nodata &   \nodata \\  
 G070.44-01.54a2 &20h14m31.88s &  31d57m05.5s &     0.11 &   0.47(0.07)  &    2.30(0.22) &  11.18(0.22) &  4.61(0.51) &  &  0.09 &   0.33(0.82) &  0.41(1.04) &   9.53(0.19) &  1.67(0.50) &   0.10(0.16) \\  
 G070.72-00.63a1 &20h11m59.82s &  32d42m02.8s &     0.03 &   0.11(0.03)  &    0.38(0.06) &  13.06(0.27) &  3.35(0.67) &  &  \nodata &   \nodata &  \nodata &   \nodata &  \nodata &   \nodata \\
\enddata
    \end{deluxetable}

\begin{deluxetable}{lrrrrrrrrrrrrr}
    \tabletypesize{\footnotesize}
    \tablewidth{290pt}
    \tablecolumns{13}
    \setlength{\tabcolsep}{0.0001in}
    \tablecaption{Derived Parameters \label{tb-derived}}
    \tablehead{
    \colhead{Designation}  & \colhead{$R_\mathrm{gal}$}  &\colhead{Dist}  & \colhead{$\tau_\mathrm{^{13}CO}$} & \colhead{$\tau_\mathrm{C^{18}O}$} & \colhead{$\tau_\mathrm{HCO^+}$}   & \colhead{$T_{ex}^\mathrm{CO}$} & \colhead{$N_\mathrm{^{13}CO}$} & \colhead{$N_\mathrm{C^{18}O}$} & \colhead{$N_\mathrm{HCO^+}$} & \colhead{$N_\mathrm{HCN}$} &     \colhead{$N_\mathrm{H_2}$} &  \colhead{$X_\mathrm{HCO^+}$} & \colhead{$X_\mathrm{HCN}$}  \\
    & \colhead{(kpc)}  &\colhead{kpc}  &  & &  & \colhead{(K)} & \colhead{($10^{15}$ cm$^{-2}$)} & \colhead{($10^{15}$ cm$^{-2}$)} & \colhead{($10^{12}$ cm$^{-2}$)} & \colhead{($10^{12}$ cm$^{-2}$)} &     \colhead{($10^{21}$ cm$^{-2}$)}   & \colhead{($10^{-11}$)} & \colhead{($10^{-11}$)}
    }
    \startdata
   G026.45+08.02a1  &  8.10 &   0.45 &   1.79 &   0.40  &  0.58 &  11.01(0.56) &   14.71(1.23) &  1.81(0.31) &  0.74(0.14) &   0.62(0.92) &  10.37(1.79) &   7.11(1.85) &  5.95(8.95)  \\ 
 G031.26-05.37a1  &  8.61 &  14.67 &   0.82 &   \nodata  &  0.35 &  11.70(0.55) &    6.90(0.52) &  \nodata &  0.46(0.17) &   \nodata &   5.39(0.41) &   8.50(3.28) &  \nodata  \\ 
 G038.36-00.95a1  &  7.60 &   1.20 &   1.32 &   0.16  &  0.12 &  10.73(0.31) &   21.45(0.95) &  2.50(0.29) &  0.52(0.12) &   \nodata &  12.64(1.48) &   4.11(1.06) &  \nodata  \\ 
 G043.02+08.36a1  &  8.39 &   0.15 &   0.57 &   \nodata  &  1.32 &  11.66(0.57) &    4.19(0.34) &  \nodata &  1.14(0.37) &   \nodata &   3.13(0.26) &  36.32(12.21) &  \nodata  \\ 
 G057.17+03.41a1  &  8.11 &   0.77 &   0.64 &   0.13  &  0.10 &  12.00(0.33) &   11.20(0.55) &  1.45(0.25) &  0.11(0.04) &   \nodata &   8.34(1.43) &   1.27(0.48) &  \nodata  \\ 
 G058.97-01.66b1  &  7.63 &   2.10 &   0.26 &   \nodata  &  0.18 &   9.67(0.59) &    3.77(0.60) &  \nodata &  0.16(0.05) &   \nodata &   2.41(0.38) &   6.49(2.49) &  \nodata  \\ 
 G070.44-01.54a2  &  8.17 &   1.25 &   0.73 &   0.07  &  0.42 &  16.04(0.17) &   38.37(0.72) &  2.77(0.32) &  4.05(0.82) &   1.43(3.60) &  16.13(1.86) &  25.09(5.85) &  8.88(22.36)  \\ 
 G070.72-00.63a1  &  8.04 &   2.26 &   0.57 &   0.12  &  0.08 &  14.72(0.50) &   48.86(3.07) &  7.35(1.38) &  0.57(0.14) &   \nodata &  41.57(7.81) &   1.37(0.42) &  \nodata  \\
\enddata
    \end{deluxetable}

\clearpage
\end{landscape}
\clearpage

\begin{deluxetable}{lrrrrrrrrr}
    \tabletypesize{\footnotesize}
    \tablewidth{18cm}
    \tablecolumns{10}
    \tablecaption{Statistics of Derived Parameters \label{tb-derived-stat}}
    \tablehead{
       & \colhead{$N_\mathrm{^{13}CO}$} & \colhead{$N_\mathrm{C^{18}O}$} & \colhead{$N_\mathrm{HCO^+}$} & \colhead{$N_\mathrm{HCN}$} &     \colhead{$N_\mathrm{H_2}$} & \colhead{$X_\mathrm{^{13}CO}/X_\mathrm{C^{18}O}$} & \colhead{$X_\mathrm{HCO^+}$} & \colhead{$X_\mathrm{HCN}$} & \colhead{$X_\mathrm{HCO^+}/X_\mathrm{HCN}$} \\
     & \colhead{($10^{15}$ cm$^{-2}$)} & \colhead{($10^{15}$ cm$^{-2}$)} & \colhead{($10^{12}$ cm$^{-2}$)} & \colhead{($10^{12}$ cm$^{-2}$)} &     \colhead{($10^{21}$ cm$^{-2}$)} &  & \colhead{($10^{-11}$)} & \colhead{($10^{-11}$)} &
    }
    \startdata
Min.   &   3.77 &  0.22 &  0.07 &  0.24 &   1.45 &   3.40 &   0.30 &  1.26 & 0.05 \\
Max.   &  56.63 &  8.68 & 24.16 & 16.15 &  70.90 &  36.56 & 152.15 & 55.81 & 5.07 \\
Median &  11.95 &  0.94 &  0.76 &  0.62 &   8.93 &  12.49 &   8.51 & 20.55 & 1.84 \\
Mean   &  15.27 &  1.79 &  1.69 &  1.57 &  12.76 &  11.45 &  14.86 & 13.56 & 1.40 \\
STD    &   9.41 &  1.42 &  2.73 &  1.91 &   9.82 &   5.33 &  18.79 & 10.90 & 0.99 
    \enddata
\end{deluxetable}

\begin{deluxetable}{lrrrr}
    \tabletypesize{\footnotesize}
    \tablewidth{9cm}
    \tablecolumns{5}
    \tablecaption{Statistics of Some Parameters of PGCCs  \label{tb-pgcc-par}}
    \tablehead{
       & \colhead{$N_\mathrm{H_2}$} & \colhead{$T_\mathrm{dust}$} & \colhead{$M_\mathrm{clump}$} & \colhead{$L_\mathrm{clump}$} \\
     & \colhead{($10^{20}$ cm$^{-2}$)} & \colhead{(K)}  & \colhead{(\msun)} & \colhead{(\lsun)} 
    }
    \startdata
    \multicolumn{5}{l}{All PGCCs}\\
   \hline
   Min.    &    0.07 &   8.55 &                  0.05 &             0.01 \\
   Max.    & 1189.23 &  30.00 &  $1.26\times10^{5}$   & $2.92\times10^6$ \\
   Median  &    4.12 &  14.53 &                  4.55 &             2.56 \\
   Mean    &   10.72 &  14.60 &                797.29 &          2299.61 \\
   STD     &   25.58 &   1.98 &               4771.02 &         52018.39 \\
   \hline
   \multicolumn{5}{l}{PGCCs detected in dense gas tracers}\\
   \hline
   Min.    &    2.52 &   9.20 &                  0.80 &             0.37 \\
   Max.    &  136.95 &  14.52 &               4428.39 &          1567.50 \\
   Median  &   24.53 &  12.35 &                 14.31 &             4.33 \\
   Mean    &   32.28 &  12.22 &                121.76 &            42.65 \\
   STD     &   25.21 &   1.13 &                532.11 &           188.13 \\
   \hline
   \multicolumn{5}{l}{PGCCs not detected in dense gas tracers}\\
   \hline
   Min.    &    2.65 &   9.20 &                  0.54 &             0.30 \\
   Max.    &  156.91 &  17.31 &               3650.42 &           756.06 \\
   Median  &   15.30 &  12.29 &                  8.04 &             3.03 \\
   Mean    &   20.72 &  12.32 &                176.56 &            38.31 \\
   STD     &   21.13 &   1.27 &                713.28 &           147.51 
    \enddata
    \tablecomments{All the parameters are from the PGCC catalog \citep{pla15}.}
    \end{deluxetable}

\end{CJK*}
\end{document}